%% file: main.tex
\documentclass[lettersize,journal]{IEEEtran}
\usepackage{amsmath,amsfonts}
\usepackage{algorithmic}
\usepackage{array}
\usepackage{array}
\newcolumntype{C}[1]{>{\centering\arraybackslash}p{#1}}
\usepackage{textcomp}
\usepackage{stfloats}
\usepackage{url}
\usepackage{verbatim}
\usepackage{graphicx}
\usepackage{booktabs} 
\usepackage{multirow}
\usepackage{makecell}
\usepackage{wrapfig}
\usepackage{tabularx}
\usepackage{adjustbox}
\usepackage{hyperref} 
\usepackage{xspace}
\usepackage{amssymb}
\usepackage{amsmath}
\usepackage{amsfonts}
\usepackage{wrapfig}
\usepackage{algorithm}
\usepackage{algorithmic}
\usepackage{pifont}
\usepackage{color}
\usepackage{colortbl}
\usepackage{babel}
\usepackage{graphicx}
\usepackage{subcaption}
\def\BibTeX{{\rm B\kern-.05em{\sc i\kern-.025em b}\kern-.08em
    T\kern-.1667em\lower.7ex\hbox{E}\kern-.125emX}}
\usepackage{balance}
\def\pzo{\phantom{0}}
\begin{document}
\title{RWKV-UNet: Improving UNet with Long-Range Cooperation for \\Effective Medical Image Segmentation }
\author{Juntao Jiang, Jiangning Zhang,  Weixuan Liu, Muxuan Gao, Xiaobin Hu, Zhucun Xue, Yong Liu, Shuicheng Yan\\ \tt Code: \href{https://github.com/juntaoJianggavin/RWKV-UNet}{\textcolor{magenta}{https://github.com/juntaoJianggavin/RWKV-UNet}}}

\maketitle
\input{secs/0_abstract}    
\input{secs/1_introduction}
\input{secs/2_related_work}
\input{secs/3_method}

\input{secs/4_experiment}
\input{secs/5_conclusion}

{
    \small
    \bibliographystyle{IEEEtran}
    \bibliography{main}
}


\end{document}

%% file: secs/0_abstract.tex
\begin{abstract}
In recent years, significant advancements have been made in deep learning for medical image segmentation, particularly with convolutional neural networks (CNNs) and transformer models. However, CNNs face limitations in capturing long-range dependencies, while transformers suffer from high computational complexity. To address this, we propose RWKV-UNet, a novel model that integrates the RWKV (Receptance Weighted Key Value) structure into the U-Net architecture. This integration enhances the model's ability to capture long-range dependencies and to improve contextual understanding, which is crucial for accurate medical image segmentation. We build a strong encoder with developed Global-Local Spatial Perception (GLSP) blocks combining CNNs and RWKVs. We also propose a Cross-Channel Mix (CCM) module to improve skip connections with multi-scale feature fusion, achieving global channel information integration. Experiments on 11 benchmark datasets show that the RWKV-UNet achieves state-of-the-art performance on various types of medical image segmentation tasks. Additionally, smaller variants, RWKV-UNet-S and RWKV-UNet-T, balance accuracy and computational efficiency, making them suitable for broader clinical applications.
\end{abstract}

%% file: secs/1_introduction.tex
\section{Introduction} \label{sec:introduction}

Computer-aided medical image analysis is crucial in modern healthcare, where deep learning–driven automated segmentation precisely delineates anatomical structures and lesions, thereby enhancing visualization, improving diagnostic accuracy, and supporting personalized treatment planning. U-Net~\cite{ronneberger2015u} is a seminal segmentation architecture featuring an encoder-decoder structure with skip connections, allowing it to capture both high-level semantic context and fine-grained details, which has inspired numerous improved variants.

Medical images are typically high-resolution and exhibit complex anatomical structures with diverse lesion appearances, posing challenges for segmentation tasks that demand both fine local detail and coherent global context. Local features are essential for accurate boundary delineation and small lesion detection, whereas long-range dependencies ensure structural consistency and cross-slice contextual integration. CNN-based designs excel at local feature extraction through convolutional kernels but struggle to model long-range dependencies. Transformers-based ~\cite{vaswani2017attention, dosovitskiy2020image}  UNet variants like ~\cite{hatamizadeh2021swin, huang2022missformer} address this via patch-based image processing and self-attention, improving segmentation accuracy, but the $O(N^2)$ computational cost impacts efficiency, especially when processing high-resolution images.
\begin{figure}[t]
  \centering
  \includegraphics[width=\linewidth]{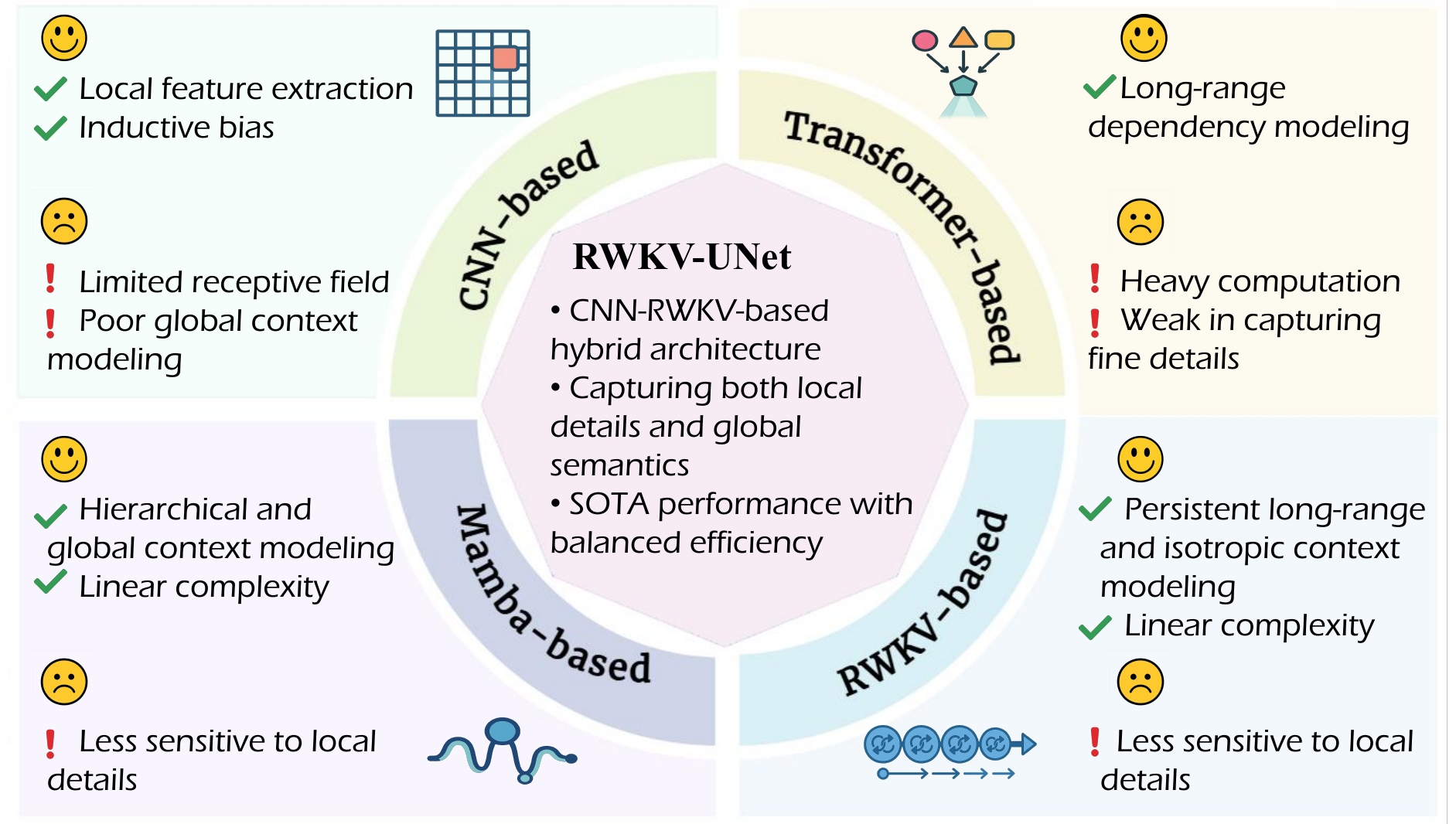}
  \caption{Comparative analysis of CNN-, Transformer-, Mamba-, RWKV-, and hybrid-based segmentation models, highlighting their respective strengths and weaknesses.}
  \label{fig:comparison_models}
\end{figure}

Recent linear-attention models such as Mamba~\cite{gu2023mamba, zhu2024vision} and RWKV~\cite{peng2023rwkv, duan2024vision} have emerged as efficient alternatives, both achieving linear complexity while maintaining long-range modeling capability, and UNet variants based on these~\cite{ruan2024vm,ma2024u,chen2025zig} have been developed for medical image segmentation. Mamba leverages state-space dynamics and selective scanning for efficient global information propagation, whereas RWKV introduces a gated recurrent mechanism that combines recurrence inductive bias with Transformer-style context aggregation. From a modeling perspective, RWKV’s gated accumulation enables more persistent and isotropic context propagation capability, yielding broader effective receptive fields and more stable long-range dependency modeling in medical image segmentation. Integrating CNNs with these global modeling mechanisms unifies precise local feature extraction and efficient global context understanding, addressing the inherent limitations of either CNNs or Transformers alone. Building on RWKV’s advantages, CNN–RWKV hybrids are expected to provide more balanced representations, offering a promising trade-off between segmentation accuracy and computational efficiency. Fig. \ref{fig:comparison_models} compares the strengths and weaknesses of each model.

Based on the advantages discussed above, we propose a novel RWKV-UNet for medical image segmentation, explicitly integrating convolutional layers with the spatial mixing capabilities of RWKV. This design leverages RWKV’s strength in capturing long-range dependencies and directionally consistent context propagation while retaining the precise local feature extraction of convolution, aiming to produce more accurate and context-aware segmentation results. We evaluate our model on 11 benchmark datasets, where it achieves state-of-the-art (SOTA) performance, demonstrating its effectiveness and potential as a robust tool for medical image analysis.


In summary, our contributions are as follows:
\begin{itemize} 
\item We propose an innovative RWKV-based Global-Local Spatial Perception (GLSP) module to build encoders for medical image segmentation. This approach efficiently integrates global and local information, and its feature extraction capabilities are augmented via pre-training.

\item  We design a Cross-Channel Mix (CCM) module to improve the skip connections in U-Net, which facilitates the fusion of multi-scale information and enhances the feature representation across different scales.

\item Our RWKV-UNet demonstrates superiority and efficiency, achieving SOTA segmentation performance on 11 datasets of different imaging modalities. Additionally, the relatively compact models, namely RWKV-UNet-S and RWKV-UNet-T, strike a balance between performance and computational efficiency, making them adaptable to a wider range of application scenarios. \end{itemize}

%% file: secs/2_related_work.tex
\section{Related Work} \label{sec:related_work}
\subsection{Architectures for Medical Image Segmentation}
U-Net~\cite{ronneberger2015u} is a deep learning architecture for biomedical image segmentation, featuring a U-shaped encoder-decoder structure with skip connections that preserve spatial details while extracting and restoring features. Variations based on U-Net improve its performance through various means, which include using developed encoders ~\cite{zhang2018road, jiang2024lv}, gating mechanisms to focus on important features~\cite{oktay2018attention}, improving skip connection formats~\cite{zhou2018unet++,huang2020unet, qian2024multi} to recover more spatial details, as well as adaptations for 3D images~\cite{cciccek20163d,milletari2016v}.  Some studies have explored combining MLP/KAN with UNet to capture long-range dependencies~\cite{valanarasu2022unext,liu2024rolling,li2024ukanmakesstrongbackbone}. nnU-Net~\cite{isensee2021nnu} stands out as a self-configuring framework that automatically adapts preprocessing, architecture, and training settings to different medical datasets, serving as a robust baseline in many clinical segmentation challenges. These improvements allow U-Net to demonstrate higher accuracy and adaptability in various image segmentation tasks. 

\noindent \textbf{Attention-based Improvements.} Attention-based methods have demonstrated strong performance in visual tasks due to their ability to capture global context. The self-attention mechanism of Transformers allows models to focus on the most informative regions, which is crucial in medical image segmentation for precise delineation of anatomical structures and lesions. Pure Transformer U-Nets~\cite{hatamizadeh2021swin, huang2022missformer, azad2022transdeeplab, chen2023ms,10598359} rely entirely on self-attention, enabling effective long-range dependency modeling but often at high computational cost. Hybrid CNN-Transformer architectures~\cite{chen2021transunet, wang2022mixed, wang2022uctransnet, heidari2023hiformer, Kang_2024_WACV} combine CNNs’ local feature extraction with Transformers’ global modeling, preserving fine structural details while capturing complex tissue and lesion relationships. More recently, linear-attention sequence models such as Mamba~\cite{gu2023mamba, zhu2024vision} and SSM-based U-Net variants~\cite{ruan2024vm, ma2024u, wang2024mamba, liao2024lightm, ma2024semi} efficiently capture long-range dependencies with linear complexity, offering a favorable balance between segmentation accuracy and computational efficiency. And there are also attention-based attenpt for segmentation in medical videos~\cite{11125515}. Despite the progress,existing parallel global-local fusion suffers from high computational cost and can interfere with fine feature representation, while some global modeling may be unnecessary in shallow layers. Additionally, existing approaches may struggle to capture long-range dependencies stably, as they lack recurrent inductive bias that support stable context propagation.

\subsection{Receptance Weighted Key Value (RWKV) }
Receptance Weighted Key Value (RWKV)~\cite{peng2023rwkv} models sequential dependencies via a weighted combination of past key-value pairs modulated by a learnable gate, avoiding the quadratic cost of traditional attention. Adapted to vision tasks as Vision-RWKV (VRWKV)~\cite{duan2024vision}, it efficiently captures long-range spatial dependencies in high-resolution images with linear complexity and reduced memory overhead, and has also been extended to broader visual domains ~\cite{he2025pointrwkv,yin2024video}.

\noindent \textbf{RWKV for Medical Imaging.} Restore-RWKV~\cite{yang2024restorerwkvefficienteffectivemedical} applies linear Re-WKV attention with an Omni-Shift layer for efficient multi-directional information propagation in image restoration. BSBP-RWKV~\cite{zhou2024bsbp} integrates RWKV with Perona-Malik diffusion to achieve high-precision segmentation, preserving structural and pathological details. RWKVMatch~\cite{he2025rwkvmatch} leverages Vision-RWKV-based global attention, cross-fusion mechanisms, and elastic transformation integration to handle complex deformations in medical image registration. Zigzag RWKV-in-RWKV (Zig-RiR) model~\cite{chen2025zig} introduces a nested RWKV structure with Outer and Inner blocks and a Zigzag-WKV attention mechanism to capture both global and local features while preserving spatial continuity. These innovations demonstrate RWKV’s ability to capture long-range dependencies, retain fine anatomical features, and scale to high-resolution images. However, systematic exploration of hybrid architectures combining RWKV with local feature extractors remains limited, leaving opportunities to balance local detail, global context, and computational efficiency in medical image segmentation.

%% file: secs/3_method.tex
\section{Methodology: RWKV-UNet} \label{sec:method}
\subsection{Overall Architecture} 
\begin{figure*}
    \centering
    \includegraphics[width=0.9\linewidth]{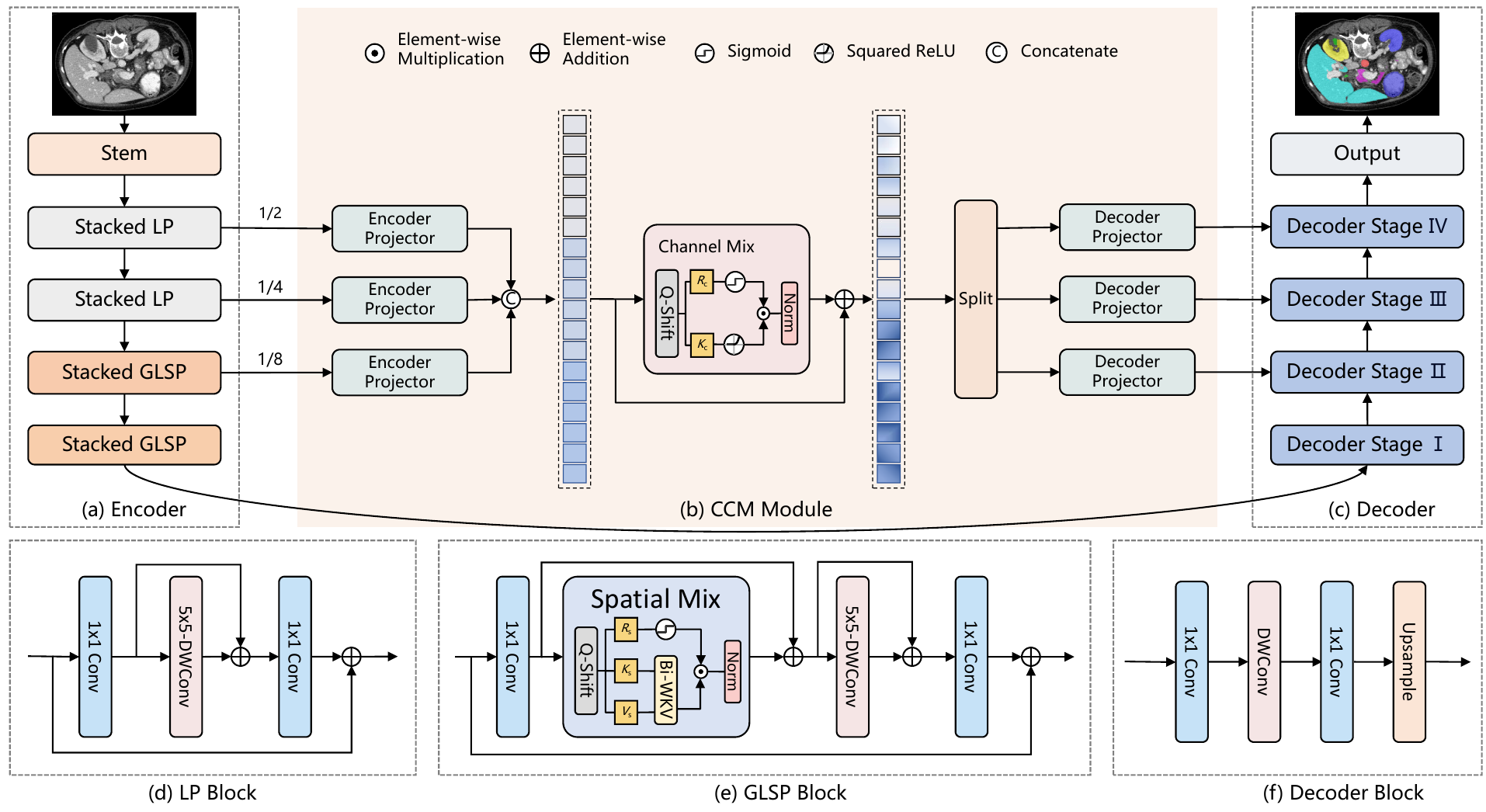}
    \caption{\textbf{Overall architecture of the proposed RWKV-UNet}. (a) the encoder with four stages constructed by stacked LP blocks and stacked GLSP blocks, (b) Cross-Channel Mix (CCM) Module for multi-scale fusion, (c) the decoder with four stages, (d) the Local Perception (LP) block, (e) the RWKV-based Global-Local Spatial Perception (GLSP) block, (f) the decoder block constructed by a point-convolution layer and a $9\times9$ \text{DW-Conv} layer, with a convolution and an upsampling operation.}
    \label{main}
\end{figure*}
The overall architecture of the proposed RWKV-UNet is presented in Fig. \ref{main}. RWKV-UNet consists of an encoder with stacked LP and GLSP blocks, a decoder, and skip connections with a Cross-Channel Mix (CCM) Module.
\subsection{Effective Encoder in RWKV-UNet} 

\paragraph{Global-Local Spatial Perception (GLSP) Block} We employ the RWKV-Spatial-Mix module and Depth-Wise Convolution (\text{DW-Conv}) with a skip connection to combine global and local dependency.  The process of expanding and reducing dimensions enhances feature representation by capturing richer details while preventing information bottlenecks. Given a feature map \( X \in \mathbb{R}^{C_{\mathrm{in}} \times H \times W} \), the processing flow through the GLSP module is as follows: Normalize and project the input \( X \) to an intermediate dimension $C_{mid}$ using a $1\times1$ \text{Convolution} layer:

\begin{equation}
I_1 = \text{LayerNorm} (\text{Conv}_{1\times1}(X)),
\end{equation}
where \(I_1 \in \mathbb{R}^{C_{\mathrm{mid}} \times H \times W} \). $C_{\mathrm{mid}}$ should be greater than $C_{\mathrm{in}}$ to achieve dimension expansion.

Divide the feature map into patches of size $1\times1$. Then the Spatial Mix in VRWKV is applied:
\begin{equation}
I_2 = \text{Unfolding}(I_1) ,
\end{equation}
\begin{equation}
I_3 = \text{SpatialMix}(\text{LayerNorm}(I_2)) + I_2 ,
\end{equation}
where $I_2 \in \mathbb{R}^{C_{\mathrm{mid}}\times N}$, $I_3 \in \mathbb{R}^{C_{\mathrm{mid}}\times N}$.

Convert the feature sequence back into a 2D feature map $I_4  = \text{Folding}(I_3) \in \mathbb{R}^{ C_{\mathrm{mid}} \times H \times W}$,  and use a $5\times5$ DW-Conv layer for local feature aggregation:
\begin{equation}
I_5 = \text{DW-Conv}(I_4) + I_4,
\end{equation}
where  \(I_5 \in \mathbb{R}^{C_{\mathrm{mid}} \times H^{\prime} \times W^{\prime}} \). \(H^{\prime} \times W^{\prime}\) is determined by the stride of \text{DW-Conv}. 

Finally, project \( I_5 \) to the output dimension and add a global skip connection:

\begin{equation}
F = \text{Conv}_{1\times1}(I_5) + X,
\end{equation}
where \(F \in \mathbb{R}^{C_{\mathrm{out}} \times H^{\prime} \times W^{\prime}} \). 

\paragraph{Local Perception (LP) Block} An LP block contains a point convolution layer, a \text{DW-Conv} layer and a point convolution layer with local and global residual skips, removing the spatial mix as well as the unfolding and folding processes from the GLSP.

\paragraph{Effective Encoder with Stacked LPs and GLSPs} We construct the RWKV-UNet encoder by stacking the LP  and GLSP blocks. The encoder comprises a stem stage followed by four stages. The first blocks in stages   \uppercase\expandafter{\romannumeral 1},   \uppercase\expandafter{\romannumeral 2},  \uppercase\expandafter{\romannumeral 3} and   \uppercase\expandafter{\romannumeral 4} do not have residual connections and use a \text{DW-Conv} with a stride of 2 to achieve downsampling. After the first block, Stages  \uppercase\expandafter{\romannumeral 1} and   \uppercase\expandafter{\romannumeral 2} are composed of stacked LP blocks without Spatial Mix. In Stages   \uppercase\expandafter{\romannumeral 3} and   \uppercase\expandafter{\romannumeral 4}, a series of GLSP blocks are stacked following the first block.

\paragraph{Scale Up} As shown in Table \ref{rwkv_unet_encoder_stages}, we implement three different sizes of encoders with different numbers of LP and GLSP blocks in each stage (depths), different embedding dimensions and expansion ratios for $C_{\mathrm{mid}}$ : Encoder-Tiny (Enc-T), Encoder-Small (Enc-S), and Encoder-Base (Enc-B), allowing flexibility to scale up the model according to different resource constraints and application needs. 
\paragraph{Pre-training Manner}
We pre-train the encoders of different sizes on ImageNet-1K~\cite{deng2009imagenet} with a batch size of 1024 for 300 epochs using the AdamW~\cite{loshchilov2017decoupled} optimizer. These pre-trained weights of different sizes will be used in medical segmentation tasks, enabling improved feature extraction and faster convergence.

\begin{figure}[htbp]
    \centering
    \begin{subfigure}[t]{0.24\linewidth}
        \centering
        \includegraphics[width=\linewidth]{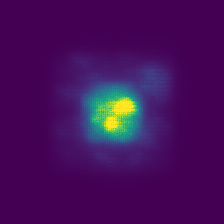}
        \caption{No Attention}
    \end{subfigure}
    \hfill
    \begin{subfigure}[t]{0.24\linewidth}
        \centering
        \includegraphics[width=\linewidth]{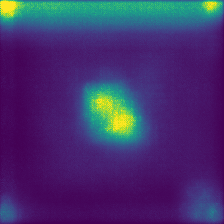}
        \caption{MHSA}
    \end{subfigure}
    \hfill
    \begin{subfigure}[t]{0.24\linewidth}
        \centering
        \includegraphics[width=\linewidth]{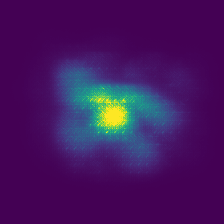}
        \caption{Bi-Mamba}
    \end{subfigure}
    \hfill
    \begin{subfigure}[t]{0.24\linewidth}
        \centering
        \includegraphics[width=\linewidth]{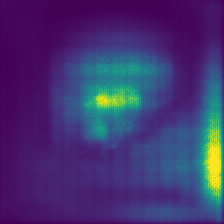}
        \caption{Spatial Mix}
    \end{subfigure}
    
    \caption{Comparison visualization of effective receptive fields of the last layer output using different attention mechanisms in the GLSP module.}
    \label{fig:glsp_erf}
\end{figure}

\begin{table}[t]
\caption{RWKV-UNet Encoder Configurations Across Stages}
\centering
\begin{tabular}{llccc}
\toprule
\textbf{Stage} & \textbf{Parameter} & \textbf{Enc-T} & \textbf{Enc-S} & \textbf{Enc-B} \\
\midrule
\textbf{Stem}& Dimension & 24 & 24 & 24 \\
\midrule
\multirow{4}{*}{\textbf{Stage   \uppercase\expandafter{\romannumeral 1}}} & Depth & 2 & 3 & 3 \\
& Dimension  & 32 & 32 & 48 \\
& Expansion & 2.0 & 2.0 & 2.0 \\
& Spatial Mix & \ding{55} & \ding{55} & \ding{55} \\
\midrule
\multirow{4}{*}{\textbf{Stage   \uppercase\expandafter{\romannumeral 2}}} & Depth & 2 & 3 & 3 \\
& Dimension   & 48 & 64 & 72 \\
& Expansion  & 2.5 & 2.5 & 2.5 \\
& Spatial Mix & \ding{55} & \ding{55} & \ding{55} \\
\midrule
\multirow{4}{*}{\textbf{Stage \  \uppercase\expandafter{\romannumeral 3}}} & Depth & 4 & 6 & 6 \\
& Dimension  & 96 & 128 & 144 \\
& Expansion  & 3.0 & 3.0 & 4.0 \\
& Spatial Mix & \checkmark & \checkmark & \checkmark \\
\midrule
\multirow{4}{*}{\textbf{Stage   \uppercase\expandafter{\romannumeral 4}}} & Depth & 2 & 3 & 3 \\
& Dimension  & 160 & 192 & 240 \\
& Expansion& 3.5 & 4.0 & 4.0 \\
& Spatial Mix & \checkmark & \checkmark & \checkmark \\
\midrule
\multicolumn{2}{c}{\textbf{Parameters}}   & 2.94M  & 9.42M  & 16.69M \\
\midrule
\multicolumn{2}{c}{\textbf{FLOPs}} &  1.83G & 4.73G & 8.96G \\

\bottomrule
\end{tabular}
\label{rwkv_unet_encoder_stages}
\end{table}
\begin{table}[htp]
\caption{Comparison results of models with different attention mechanisms on Synapse dataset. }
\renewcommand{\thetable}{B1}
\centering
\resizebox{\linewidth}{!}{
\begin{tabular}{cccc}
\hline 
\multirow{2}{*}{Attention Type}&\multicolumn{2}{c}{Average}  & \multirow{2}{*}{FLOPs} \\
\cline{2-3} 
&  HD95 $\downarrow$ & DSC $\uparrow$&\\ 
\hline
Multi-Head Self-Attention~\cite{vaswani2017attention}&  29.32 & 77.03&18.55G \\ 
Focused Linear Attention~\cite{han2023flatten}&    \underline{27.66}&76.89& \underline{11.18G} \\ 
  Bi-Mamba~\cite{zhu2024vision}& 29.10 & \underline{78.13} &14.53G\\ 
  
 RWKV Spatial-Mix (ours)~\cite{ruan2024vm}&  \textbf{25.01} & \textbf{78.14} & \textbf{11.11G}\\ 
\hline 
\end{tabular}
}
\label{attention2}
\end{table}
\paragraph{Comparison with other attention types} 
We conduct experiments by replacing the spatial mix in the encoder of RWKV-UNet with other attention mechanisms, such as self-attention~\cite{vaswani2017attention},  focused linear attention~\cite{han2023flatten}, and Bi-Mamba used in Vision Mamba~\cite{zhu2024vision}, as well as by removing this module entirely.  The total training epochs are 100. The results shown in Table \ref{attention2} indicate that the model with Spatial-Mix achieves the best DSC with the lowest computational cost. Comparison visualizations of effective receptive fields of the last layer output using different attention mechanisms in the GLSP module are shown in Fig. \ref{fig:glsp_erf}.

\subsection{Large Kernel-based Decoder Design}

We design our CNN-based decoder block with a point-convolution layer and a $9\times9$ \text{DW-Conv} layer, following a point operation layer and an upsampling operation. Consider the input feature map $X \in \mathbb{R}^{C_{\text {in }} \times H \times W}$; in depthwise convolution, the number of input channels is $C_{\mathrm{in}}$, and the second point convolution maps the $C_{\text {in }}$ channels to $C_{\text {out }}$.

\subsection{Cross-Channel Mix for Multi-scale Fusion} 
Inspired by the Channel Mix in VRWKV, we propose Cross-Channel Mix (CCM), a module that can effectively extract channel information across multi-scale encoder features. By capturing the rich global context along channels,  CCM further enhances the extraction of long-range contextual information.

Define output feature map of \text{Stage   \uppercase\expandafter{\romannumeral 1}}, \text{Stage   \uppercase\expandafter{\romannumeral 2}} and \text{Stage   \uppercase\expandafter{\romannumeral 3}} of the designed encoder are $F_1, F_2$, and $F_3$, which has different Spatial dimensions and channel counts: $
F_1 \in \mathbb{R}^{C_1 \times H_1 \times W_1}, \quad F_2 \in \mathbb{R}^{C_2 \times H_2 \times W_2}, \quad F_3 \in \mathbb{R}^{C_3 \times H_3 \times W_3}
$. ($H_1$ \textgreater $H_2$\textgreater $H_3$, $W_1$ \textgreater $W_2$ \textgreater $W_3$ and $C_3$ \textgreater $C_2$ \textgreater $C_1$).

We reshape and map smaller features to match the largest feature’s size and dimension:
\begin{equation}
\begin{gathered}
\tilde{F}_i=\operatorname{Conv}_{C_i}\left(\operatorname{Upsample}\left(F_i\right)\right), \quad i=2,3 \\
\tilde{F}_1=\operatorname{Conv}_{C_1}\left(F_1\right), 
\end{gathered}
\end{equation}
where $\tilde{F}_i \in \mathbb{R}^{C_1 \times H_1 \times W_1}, \quad i=1,2,3 $

We concatenate the adjusted feature maps along the channel dimension to produce a combined feature map:
\begin{equation}
F_{\text{cat}} = \text{Concat}(\tilde{F}_1, \tilde{F}_2, \tilde{F}_3), 
\end{equation}
where $F_{\text{cat}} \in \mathbb{R}^{3C_1 \times H_1 \times W_1}$.

Divide the feature map into patches: $F_{\text{unf}} = \text{Unfolding}(F_{\text{cat}})\in \mathbb{R}^{3C_1 \times N}$. Then apply the Channel Mix in VRWKV, performing multi-scale global feature fusion in the channel dimension:

\begin{equation}
F_{\text{mix}} = \text{ChannelMix}(\text{LayerNorm}(F_{\text{unf}}))+F_{\text{unf}},
\end{equation}
where  $F_{\text{mix}}\in \mathbb{R}^{3C_1 \times N_1}$. Project it back to 2D feature, $F_{\text{fold}}=\text{Folding}(F_{\text{mix}})\in \mathbb{R}^{3C_1 \times H_1 \times W_1}$, and $N_1=H_1 \times W_1$.

The folded feature map $F_{\text {fold }}$ is split back into three separate features:$[F_{\text{fold}}^{(1)}, F_{\text{fold}}^{(2)}, F_{\text{fold}}^{(3)}]$. Each split feature undergoes a reshape operation and a convolution to restore its original sizes and dimensions:
\begin{equation}
F_i^{\prime}=\operatorname{Conv}_{C_i}\left(\operatorname{Reshape}\left(F_{\text {fold }}^{(i)}\right)\right), \quad i=1,2,3
\end{equation}
where $F_1'\in \mathbb{R}^{C_1 \times H_1 \times W_1}$ , $F_2'\in \mathbb{R}^{C_2 \times H_2 \times W_2}$ and $F_3' \in \mathbb{R}^{C_3 \times H_3 \times W_3}$. $F_1'$,$F_2'$ and $F_3'$ will be concatenated with the output of the decoder \text{Stage  \uppercase\expandafter{\romannumeral 3}}, \text{Stage   \uppercase\expandafter{\romannumeral 2}} and \text{  \uppercase\expandafter{\romannumeral 1}}.

%% file: secs/4_experiment.tex
\section{Experiments}
\begin{table*}[h]
\caption{Comparison results on Synapse dataset. The evaluation metrics are HD95 (mm) and DSC in (\%). DSC are reported for individual organs. $\uparrow \downarrow$ denotes the higher (lower) the better. $-$ means missing data from the source. \textbf{Bold} and \underline{underline} represent the best and the second best results. $^*$ denotes that the experiment is conducted by us.}
\centering
\resizebox{\textwidth}{!}{ 
\begin{tabular}{lcccccccccc}
\hline 
\multirow{2}{*}{ Methods } & \multicolumn{2}{c}{ Average } & \multirow{2}{*}{ Aotra } & \multirow{2}{*}{ Gallbladder } & \multirow{2}{*}{ Kidney (Left) } & \multirow{2}{*}{ Kidney (Right) } & \multirow{2}{*}{ Liver } & \multirow{2}{*}{ Pancreas } & \multirow{2}{*}{ Spleen } & \multirow{2}{*}{ Stomach } \\
\cline{2-3} & HD95 $\downarrow$ & DSC $\uparrow$ & & & & & & & & \\
\hline 
U-Net~\cite{ronneberger2015u} &39.70& 76.85 &\underline{89.07}& 69.72 &77.77 &68.60 &93.43& 53.98& 86.67 &75.58\\
R50 U-Net~\cite{he2016deep,ronneberger2015u} &36.87 &74.68& 87.74& 63.66 &80.60 &78.19 &93.74 &56.90 &85.87 &74.16\\
ViT~\cite{dosovitskiy2020image} + CUP~\cite{chen2021transunet} & 36.11 & 67.86 & 70.19 & 45.10 & 74.70 & 67.40 & 91.32 & 42.00 & 81.75 & 70.44 \\
TransUNet~\cite{chen2021transunet} & 31.69 & 77.48 & 87.23 & 63.16 & 81.87 & 77.02 & 94.08 & 55.86 & 85.08 & 75.62 \\
SwinUNet~\cite{cao2022swin} & 21.55 & 79.13 & 85.47 & 66.53 & 83.28 & 79.61 & 94.29 & 56.58 & 90.66& 76.60 \\
TransClaw U-Net~\cite{chang2021transclaw} & 26.38 & 78.09 & 85.87 & 61.38 & 84.83 & 79.36 & 94.28 & 57.65 & 87.74 & 73.55 \\
MT-UNet~\cite{wang2022mixed} &  26.59& 78.59& 87.92 &64.99& 81.47& 77.29 &93.06& 59.46 &87.75& 76.81\\
UCTransNet~\cite{wang2022uctransnet}  &26.75& 78.23 &  - &-& - &- &-& - &-& -\\
MISSFormer~\cite{huang2022missformer} & 18.20 & 81.96 & 86.99 & 68.65 & 85.21 & 82.00 & 94.41 & 65.67 &91.92 & 80.81 \\
TransDeepLab \cite{azad2022transdeeplab}  &21.25&80.16&  86.04 &69.16& 84.08 &79.88 &93.53& 61.19 &89.00& 78.40\\
LeVit-Unet-384~\cite{xu2023levit} &16.84& 78.53& 87.33 &62.23& 84.61 &80.25& 93.11& 59.07 &88.86 &72.76 \\
MS-UNet~\cite{chen2023ms} & 18.97&80.44& 85.80& 69.40 &85.86 & 81.66 &94.24 &57.66 &90.53 &78.33\\
HiFormer-L~\cite{heidari2023hiformer}&19.14 &80.69& 87.03 &68.61& 84.23& 78.37 &94.07& 60.77 &90.44 &82.03\\
PVT-CASCADE~\cite{rahman2023medical}  &20.23& 81.06& 83.01 &\textbf{70.59}& 82.23 &80.37 &94.08 &64.43& 90.10 &83.69\\
TransCASCADE~\cite{rahman2023medical} & 17.34&82.68&  86.63 &68.48& 87.66& \textbf{84.56} &94.43 &65.33 &90.79& 83.52\\
MetaSeg-B~\cite{Kang_2024_WACV} &-& 82.78& - &-& - &- &-&- &-& -\\
VM-UNet~\cite{ruan2024vm}& 19.21 &81.08 & 86.40 &69.41& 86.16 &82.76& 94.17& 58.80 &89.51 &81.40 \\
Hc-Mamba~\cite{10944508} & 26.34 & 79.58 &\textbf{89.93}& 67.65 &84.57 &78.27 &95.38 &52.08 &89.49 &79.84 \\
Mamba-UNet$^*$~\cite{wang2024mamba}&17.32&73.36 &81.65 &55.39  &81.86&72.76  &92.35& 45.28& 88.22& 69.35\\
PVT-EMCAD-B2~\cite{rahman2024emcad}&15.68& \underline{83.63}& 88.14& 68.87 &88.08& 84.10& 95.26&\textbf{68.51} &\underline{92.17} & \underline{83.92}\\
SelfReg-SwinUnet~\cite{zhu2024selfregunetselfregularizedunetmedical, cao2022swin}&-& 80.54& 86.07& 69.65& 85.12 &82.58 &94.18& 61.08 &87.42& 78.22\\

Zig-RiR (2D)~\cite{chen2025zig} (256$\times$256)$^*$&28.51& 74.61&82.73&64.19&79.13 &68.25 & 93.71 & 50.14& 88.46 & 70.29\\
\hline
RWKV-UNet-T (ours) & 11.89 &81.87&  88.10&68.42&88.88 &82.61  &95.15&60.52&90.23&81.09 \\
RWKV-UNet-S (ours) &\underline{11.01} &82.93&88.30&65.92& \underline{88.47}& 84.11 &\underline{95.45}&66.51&91.51&83.17 \\
RWKV-UNet (ours) & \pzo\textbf{8.85}& \textbf{84.29}& 88.71 &\underline{70.39}& \textbf{89.58}& \underline{84.36} &\textbf{95.70}&\underline{66.58}&\textbf{93.54}& \textbf{85.49}\\

\hline
\end{tabular}
}
\label{Synapse}
\end{table*}

\begin{table}[h]
\caption{Comparison results on ACDC
dataset. The evaluation metric is DSC in (\%). }
\resizebox{\linewidth}{!}{
\begin{tabular}{lcccc}
\hline Methods & Average & RV & Myo & LV \\
\hline
R50 U-Net~\cite{he2016deep,ronneberger2015u} &87.55 &87.10& 80.63 &94.92\\
ViT~\cite{dosovitskiy2020image} + CUP~\cite{chen2021transunet} & 81.45 & 81.46 & 70.71 & 92.18 \\
R50-ViT~\cite{he2016deep,dosovitskiy2020image} + CUP~\cite{chen2021transunet} & 87.57 & 86.07 & 81.88 & 94.75 \\
TransUNet~\cite{chen2021transunet} & 89.71 & 88.86 & 84.54 & 95.73 \\
SwinUNet~\cite{cao2022swin} & 90.00 & 88.55 & 85.62 & 95.83 \\
MT-UNet~\cite{wang2022mixed} &90.43 &86.64 &89.04 &95.62 \\
MS-UNet~\cite{chen2023ms} &87.74 & 85.31 &84.09 &93.82\\
MISSFormer~\cite{huang2022missformer} & 90.86 & 89.55 & 88.04 & 94.99 \\
LeViT-UNet-384s~\cite{xu2023levit} & 90.32 & 89.55 & 87.64 & 93.76 \\

PVT-CASCADE~\cite{rahman2023medical} &90.45& 87.20 &88.96& 95.19\\
VM-UNet~\cite{ruan2024vm}& 91.47&89.93 & 89.04 &95.44 \\
TransCASCADE~\cite{rahman2023medical} &91.63 &89.14 &\textbf{90.25} &95.50\\
PVT-EMCAD-B2~\cite{rahman2024emcad} & \underline{92.12} & \underline{90.65} &\underline{89.68} &96.02\\
SelfReg-SwinUnet \cite{zhu2024selfregunetselfregularizedunetmedical, cao2022swin} &91.49& 89.49 &89.27 &95.70\\
Zig-RiR (2D)~\cite{chen2025zig} (256$\times$256)$^*$ &87.42& 86.41 &81.87 &93.98\\
\hline
RWKV-UNet-T (ours) & 91.90&90.63&88.21 &\textbf{96.85}\\
RWKV-UNet-S (ours) & 91.19&89.49&87.87  &96.22 \\
RWKV-UNet (ours) & \textbf{92.29}&\textbf{91.26}& 88.78 &\underline{96.83}\\
\hline
\end{tabular}}
\label{acdc}
\end{table}

\begin{table}[h]
\centering
\caption{Comparison results on GOALS and FUGC 2025
dataset. The evaluation metric is DSC in (\%). All experiments for baselines are conducted by us. }
\resizebox{\linewidth}{!}{
\fontsize{5pt}{6pt}\selectfont

\begin{tabular}{lcccc}
\hline Methods & GOALS & FUGC 2025  \\
\hline
U-Net~\cite{ronneberger2015u} &90.99 &51.96\\
TransUNet~\cite{chen2021transunet} & 91.05 &  \underline{79.87} \\
Rolling-UNet-M~\cite{liu2024rolling} &90.84 &63.39\\
PVT-EMCAD-B2~\cite{rahman2024emcad}&90.91&57.39\\
PVTFormer~\cite{jha2024ct}&90.98&78.41\\
UKAN~\cite{li2024ukanmakesstrongbackbone} &90.59& 72.19\\
VM-UNet~\cite{ruan2024vm}&91.10& 72.94\\
H-vmunet~\cite{wu2025h}&84.18&65.99\\
Zig-RiR (2D)~\cite{chen2025zig}&83.48& 62.41 \\
\hline
RWKV-UNet-T (ours) & \underline{91.34}&78.67\\
RWKV-UNet-S (ours) & 91.13&79.43 \\
RWKV-UNet (ours) & \textbf{91.75}&\textbf{81.17}\\
\hline
\end{tabular}}
\label{oct}
\end{table}

\begin{table*}[h]
    \label{comp}
    \caption{Comparison results on BUSI, CVC-ClinicDB, CVC-ColonDB, Kvasir-SEG, ISIC 2017, GLAS and PSVFM datasets. The evaluation metric are  DSC in (\%). All experiments for baselines are conducted by us. Results with error bars are averaged over three runs with different splits. }
\setlength\tabcolsep{12.0pt}
    \centering
    \resizebox{1.0\textwidth}{!}{
    \begin{tabular}{cccccccc|ccc}
    \hline
 Methods &BUSI &CVC-ClinicDB  &CVC-ColonDB&Kvasir-SEG&ISIC 2017& GLAS&PSVFM&Params.& FLOPs \\
 \hline 
U-Net \cite{ronneberger2015u} & 74.90$\pm$1.08&  89.22$\pm$3.67&83.43$\pm$4.79&84.94$\pm$2.43 &82.94& 88.56$\pm$0.63 &68.98&\pzo7.77M &12.16G\\
R50-UNet~\cite{he2016deep, ronneberger2015u} & 78.46$\pm$1.52& 93.60$\pm$0.85& 89.53$\pm$2.99 & 90.13$\pm$1.16&  84.78& 90.64$\pm$0.23&73.94&28.78M &\pzo9.72G\\
Att-UNet \cite{oktay2018attention}  & 78.14$\pm$1.08&  90.86$\pm$3.58&84.75$\pm$6.48&86.21$\pm$1.93 &83.96& 89.14$\pm$0.48&60.61&\pzo8.73M &16.78G\\
UNet++ \cite{zhou2018unet++}& 75.74$\pm$1.60&  90.07$\pm$4.68&84.75$\pm$3.92&85.86$\pm$1.41&82.27& 88.93$\pm$0.60&71.69&\pzo9.16M &34.71G\\
TransUNet \cite{chen2021transunet} & 79.35$\pm$2.18&  93.91$\pm$2.20&90.72$\pm$3.50&91.34$\pm$0.86 &\underline{86.09}& 91.58$\pm$0.15&74.86& 105.3M&33.42G\\
FCBFormer~\cite{sanderson2022fcn}& 80.66$\pm$0.26&  91.15$\pm$3.29&85.23$\pm$4.57&91.09$\pm$0.83 &84.56& 91.13$\pm$0.14&75.23&51.96M&41.22G\\
UNext~\cite{valanarasu2022unext}& 73.19$\pm$1.94&  83.48$\pm$2.78&78.98$\pm$1.86&79.99$\pm$2.84 &84.68& 84.09$\pm$0.43&56.41&\pzo1.47M&\pzo0.58G\\
PVT-CASCADE~\cite{rahman2023medical}& 78.30$\pm$2.75&  93.45$\pm$1.44&90.97$\pm$2.03&\underline{91.40$\pm$1.12} &84.77& 91.85$\pm$0.34&71.82&35.27M &\pzo8.20G\\
Rolling-UNet-M~\cite{liu2024rolling} & 77.49$\pm$2.20&  91.91$\pm$2.49&85.20$\pm$5.20&87.41$\pm$1.74 &84.42&88.12$\pm$0.57&68.93&  \pzo7.10M&\pzo8.31G\\
PVT-EMCAD-B2~\cite{rahman2024emcad} & 79.22$\pm$0.93&  93.39$\pm$1.42&91.27$\pm$1.63&90.37$\pm$1.04 &84.05&91.77$\pm$0.08 &71.29& 26.76M&\pzo5.64G\\
U-KAN~\cite{li2024ukanmakesstrongbackbone} & 76.95$\pm$1.48&  89.88$\pm$2.82&84.66$\pm$4.62&85.78$\pm$1.46 &83.46& 87.42$\pm$0.25&66.25& 25.36M&\pzo8.08G\\
\hline
RWKV-UNet-T (ours)& 79.47$\pm$1.19&  95.05$\pm$0.64&89.56$\pm$5.10&91.26$\pm$2.01 &85.59&91.96$\pm$0.30&75.23& \pzo3.15M& \pzo3.70G\\
RWKV-UNet-S (ours)& \underline{80.92$\pm$1.51}&  \textbf{95.58$\pm$0.63}&\underline{91.30$\pm$2.79}&\textbf{91.78$\pm$1.19}&\textbf{86.38}&\underline{92.11$\pm$0.14}&\textbf{78.14} &\pzo9.70M&\pzo7.64G \\
RWKV-UNet (ours)& \textbf{81.92$\pm$0.74}&  \underline{95.26$\pm$0.61}&\textbf{92.27$\pm$2.17}&91.26$\pm$1.15 &85.32&\textbf{92.35$\pm$0.07}& \underline{76.39}&17.13M &14.50G\\

\hline
\end{tabular}}
\label{binary}
\end{table*}

\begin{figure}[htbp]
    \centering
    \includegraphics[width=\linewidth]{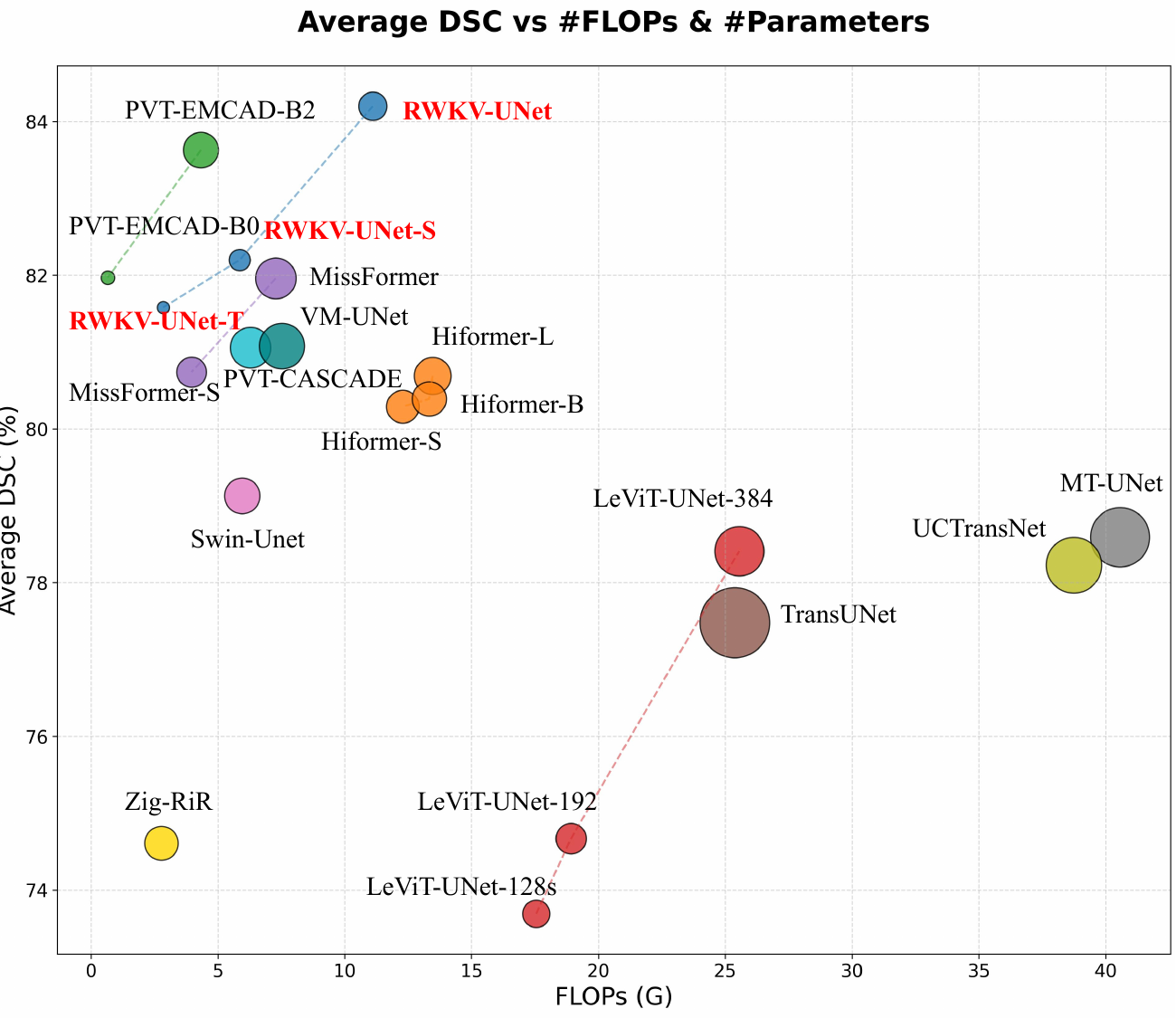}
    \caption{Performance of different methods on the Synapse multi-organ segmentation dataset.
    The average DSC (\%) is plotted against FLOPs (G). The size of each circle represents the model's parameter count. RWKV-UNet achieves SOTA performance with balanced computation cost, while RWKV-UNet-S and RWKV-UNet-T also achieve remarkable results.}
    \label{fig:FLOPS}
\end{figure}

\subsection{Experimental Setup}
\paragraph{Datasets and Metrics} 
Experiments on conducted on Synapse~\cite{landman2015miccai} for multi-organ Segmentation in CT images, ACDC~\cite{bernard2018deep} for cardiac segmentation in MRI images, GOALS~\cite{fang2022dataset} for OCT layer segmentation, BUSI~\cite{al2020dataset} for breast tumor segmentation in ultrasound images, CVC-ClinicDB~\cite{bernal2015wm}, CVC-ColonDB~\cite{bernal2012towards}, Kvasir-SEG~\cite{jha2020kvasir} for poly segmentation in endoscopy images, ISIC 2017~\cite{codella2018skin} for skin lesion segmentation in dermoscopic images, GLAS~\cite{sirinukunwattana2017gland} for gland segmentation in microscopy images, PSVFM~\cite{bano2020deep} for placental vessel segmentation in fetoscopy images and FUGC~\cite{bai202516893174} for semi-supervised cervix segmentation in ultrasound images. The average Dice Similarity Coefficient (DSC) and average 95\% Hausdorff Distance (HD95) are used as evaluation metrics. \begin{itemize}
    \item Synapse uses 18 training and 12 validation CT scans, while ACDC uses 70 training, 10 validation, and 20 testing MRI scans, following settings in ~\cite{chen2021transunet}.
    \item GOALS uses 100 training, 100 validation and 100 test images following the original setting in the challenge.
    \item BUSI, CVC-ClinicDB, CVC-ColonDB, and Kvasir-SEG are split into training, validation, and testing sets with an 8:1:1 ratio, repeated with three different random seeds.
    \item ISIC 2017 uses 2000 training, 150 validation, and 600 testing images as per the original challenge setup.
    \item GlaS uses 85 training and 80 testing whole-slide images, with the training set further split into training and validation with a 9:1 ratio across three random seeds.
    \item PSVFM uses a video-based split: Videos 1--3 for training, Video 4 for validation, and Videos 5--6 for testing, reflecting real-case variability in fetoscopic surgeries.
    \item Building on \cite{jieyunbai} and \cite{jieyunbai}, FUGC uses 50 labeled and 450 unlabeled images for training, 90 for validation, and 300 for testing.
\end{itemize}

\paragraph{Implementation Details} All experiments are performed on NVIDIA Tesla V100 with 32 GB memory. The resolution of the input images is resized to 224$\times$224 for Synapse and ACDC, and 256$\times$256 for binary segmentation tasks. The total training epochs for Synapse are 30, for ACDC are 150, and for binary segmentation are 300. The batch size is 24 for Synapse and ACDC (32 for RWKV-UNet-T on Synapse to avoid a CUDA memory problem), and 8 for GOALS and binary segmentation tasks. The initial learning rate is $1e^{-3}$ for Synapse and FUGC 2025, $5e^{-4}$ for ACDC, and $1e^{-4}$ for GOALS and binary segmentation experiments. The minimum learning rates for Synapse and ACDC are 0 and for GOALS, binary segmentation and semi-supervised experiments are 1$e^{-5}$.  The  AdamW~\cite{loshchilov2017decoupled} optimizer and CosineAnnealingLR~\cite{loshchilov2016sgdr} scheduler are used.  The baseline results on GOALS and binary segmentation tasks are run by us. The loss function for supervised tasks is a mixed loss combining cross entropy (CE) loss and dice loss~\cite{milletari2016v}:
\begin{equation}
\mathcal{L}=\alpha C E(\hat{y}, y)+\beta D i c e(\hat{y}, y),
\end{equation}
where $\alpha$ and $\beta$ are 0.5 and 0.5 for Synapse, 0.2 and 0.8 for ACDC, 0.5 and 1 for GOALS and binary segmentation.  Experiments for PVT-CASCADE~\cite{rahman2023medical} and PVT-EMCAD-B2~\cite{rahman2024emcad} use deep supervision strategy~\cite{lee2015deeply}. For semi-supervised tasks, the supervised loss on labeled data remains $\mathcal{L}_{\text{sup}} = 0.5CE(\hat{y},y) + Dice(\hat{y},y)$. In addition, a consistency loss is applied on unlabeled data, computed as the mean squared error (MSE) between predictions under different augmentations after inverting the transformations. Learning rate for TransUNet and PVTFormer are $1e^{-4}$ on FUGC 2025 for better results.

\subsection{Comparison with State-of-the-arts}
\begin{figure*}
    \centering
    \includegraphics[width=0.92\linewidth]{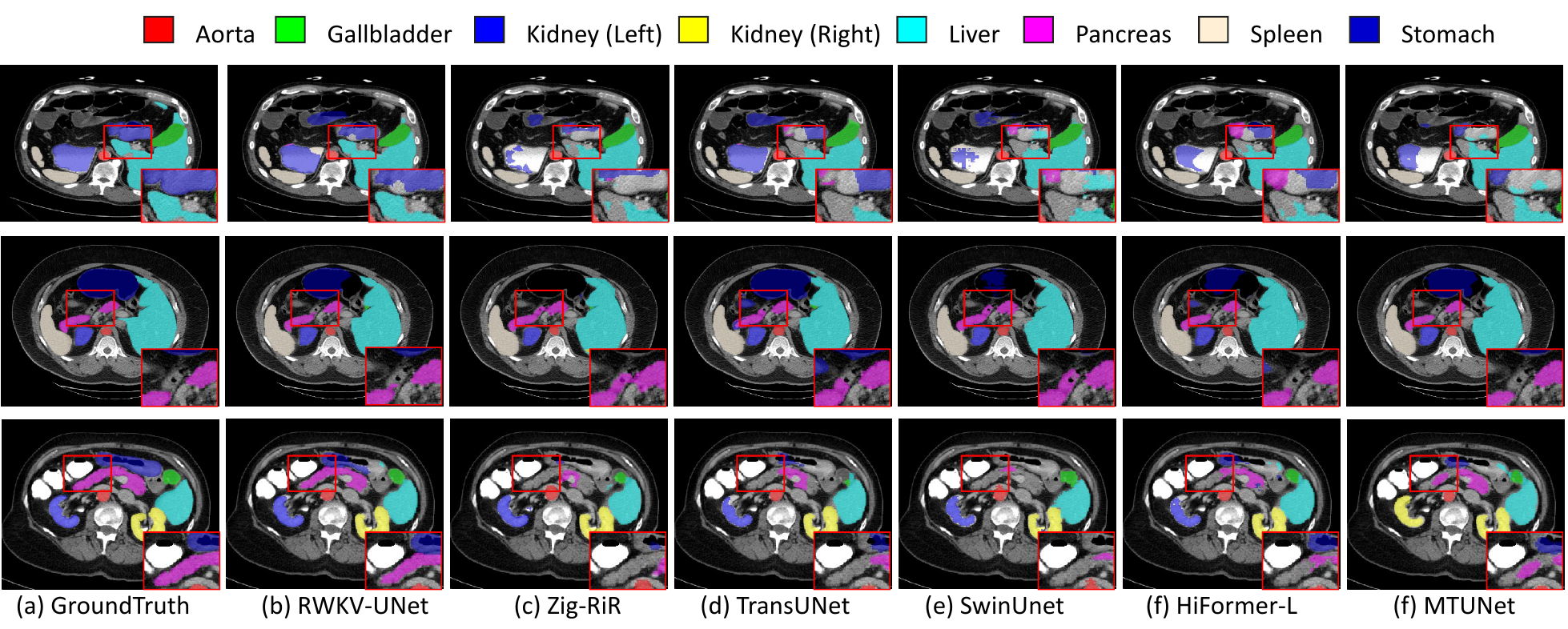}
    \caption{A qualitative comparison with previous SOTA methods on the Synapse dataset. The visual results demonstrate that our method achieves more accurate segmentation, especially in difficult tasks like pancreas segmentation. }
    \label{visual}
\end{figure*}

\paragraph{Abdomen Multi-organ Segmentation}
Table \ref{Synapse} shows that RWKV-UNet excels in abdomen organ segmentation on the Synapse dataset, achieving the highest average DSC of 84.29\% , and
surpassing all SOTA CNN-, transformer-, and Mamba-based methods. It also achieves 8.85 in HD95, which is significantly better than other models, demonstrating a strong ability to localize organ boundaries accurately.  This can be attributed to the model's ability to capture both long-range dependencies and local features. Smaller models, RWKV-UNet-T and RWKV-UNet-S, also achieve remarkable results. The DSCs with parameters and FLOPs of each model can be seen in Fig. \ref{fig:FLOPS} and visualization results can be seen in Fig. \ref{visual}.

\paragraph{Cardiac Segmentation}
Table \ref{acdc} shows that our model outperforms all SOTA models in the ACDC MRI cardiac segmentation task, achieving the best results in RV and the second-best in LV categories. The smaller models, RWKV-UNet-T and RWKV-UNet-S, also perform well.

\paragraph{OCT Layer Segmentation} 
Table \ref{oct} shows that on the small GOALS dataset, performance differences are minor and some carefully designed models even lag behind the classic U-Net, while our RWKV-UNet still achieves the highest average DSC, demonstrating its robustness and superiority. These results also reflect the model's ability to handle small datasets.
\paragraph{Binary Segmentation} Table \ref{binary} shows that the RWKV-UNet series shows remarkable performance in the tasks of breast lesion, poly, skin lesion, grand, and vessel segmentation, with much smaller parameters and computation load than TransUNet and FCBFormer. 
\paragraph{Semi-supervised Cervix Segmentation}Table \ref{oct} also shows that the RWKV-UNet achieves SOTA performance on the FUGC 2025 dataset, further indicating strong data efficiency and generalization ability, as it can fully leverage limited labeled data and mine useful information from unlabeled data to maintain high segmentation accuracy.

\subsection{Ablation Study and Additional Analysis}
\paragraph{Comparison with Different Encoders}
We replace the encoder with pre-trained weights while keeping the rest of the network architecture unchanged (dimensions vary according to the encoder). Table \ref{encoder} shows that our pretrained encoder achieves the best DSC and HD95 on the Synapse dataset, with significantly lower FLOPs.
\begin{table}[t]
\caption{Comparison results on Synapse dataset of different pretrained encoders }
\centering
\resizebox{\linewidth}{!}{
\fontsize{5pt}{6pt}\selectfont 
\begin{tabular}{cccc}
\hline 
\multirow{2}{*}{Encoder}  &\multicolumn{2}{c}{Average} &\multirow{2}{*}{FLOPs}\\
\cline{2-3} 
 & HD95 $\downarrow$ & DSC $\uparrow$\\ 
\hline 
ResNet 50~\cite{he2016deep} & \underline{9.99} &82.27&62.44 \\ 
PVT-B3-v2~\cite{wang2022pvt} & 11.47 &81.23&22.74 \\ 
PVT-B5-v2~\cite{wang2022pvt} & 10.53 &80.14&27.38 \\ 

ConvNext-small~\cite{liu2022convnet}&12.79&76.34&42.47 \\
ConvNext-base~\cite{liu2022convnet}&10.61&81.42& 74.86\\
MaxViT-small~\cite{tu2022maxvit}&10.57&\underline{83.71}&\underline{18.54} \\ 
MaxViT-base~\cite{tu2022maxvit}&11.09&82.89&29.86 \\ 
RWKV-UNet Enc-B (ours)& \pzo\textbf{8.85} & \textbf{84.29}& \textbf{11.11}\\ 

\hline 
\end{tabular}
}
\label{encoder}
\end{table}
\paragraph{Effect of Pre-training for Encoder}
Experimental results in Table \ref{pretrain}. indicate that pre-training is essential because it significantly improves feature extraction by enabling the encoder to capture intricate patterns and hierarchical information. And the inductive biases of RWKV are also not as strong as those of CNN, thus they also rely on pre-training similar to transformers. 

\begin{table}[t]
\caption{Comparison results on Synapse dataset of whether to use pre-trained weights on ImageNet-1k or not. }
\centering
\resizebox{\linewidth}{!}{
\fontsize{5pt}{6pt}\selectfont 
\begin{tabular}{cccc}
\hline 
\multirow{2}{*}{Pre-training} & \multirow{2}{*}{Training Epochs} &\multicolumn{2}{c}{Average} \\
\cline{3-4} 
& & HD95 $\downarrow$ & DSC $\uparrow$\\ 
\hline 
 w/o& 30  & 36.23 &71.66 \\ 
  w/o& 100  & 25.01 & 78.14 \\ 
 w/& 30 & \pzo\textbf{8.85} & \textbf{84.29} \\ 
\hline 
\end{tabular}
}
\label{pretrain}
\end{table}

\begin{table}[t]

\caption{Comparison results of different decoder designs for RWKV-UNet on Synapse dataset. }
\centering
\resizebox{\linewidth}{!}{ 
\fontsize{5pt}{6pt}\selectfont 
\begin{tabular}{ccccc}
\hline 
 \multirow{2}{*}{Kernel Size} & \multicolumn{2}{c}{Synapse} & {ACDC} & \multirow{2}{*}{FLOPs} \\
\cmidrule(l{2pt}r{2pt}){2-3} \cmidrule(l{2pt}r{2pt}){4-4}
& HD95 $\downarrow$ & DSC $\uparrow$ & DSC $\uparrow$ &\\ 
\hline 
 3 &  10.35 & \textbf{84.83} &\underline{92.07} &10.97G \\
 5 & \pzo\textbf{8.69} & \underline{84.67} &91.95& 11.00G \\
7 & \pzo9.97 & 83.73&91.61 & 11.05G \\
9 & \pzo\underline{8.85}& 84.29& \textbf{92.29} & 11.11G \\
11 & 11.25 & 82.67 & 92.04&11.19G \\
\hline 
\end{tabular}
}
\label{decoder}
\end{table}

\paragraph{Attention in Shallow Stages}
We evaluate the effect of retaining attention in the first two layers of RWKV-UNet-S by replacing them with spatial-mix attention and comparing against the original CNN-based shallow layers. All models are trained for 100 epochs from scratch, without pretraining. Table \ref{shallow-attention-comparison} shows negligible improvement from shallow attention on Synapse, while introducing higher computational cost, indicating that CNNs are sufficient for capturing low-level features in the shallow layers.
\begin{table}[htp]
\caption{Comparison of results on the Synapse dataset: effect of attention in shallow stages of RWKV-UNet-S}
\centering
\resizebox{\linewidth}{!}{
\fontsize{5pt}{6pt}\selectfont 
\begin{tabular}{cccc}
\hline 
\multirow{2}{*}{Shallow Attention}  & \multicolumn{2}{c}{Average}  & \multirow{2}{*}{FLOPs} \\
\cline{2-3} 
& HD95 $\downarrow$ & DSC $\uparrow$ & \\ 
\hline 
w/  & 29.43 & \textbf{76.88} & 6.91G \\ 
w/o  (ours) & \textbf{28.92} & 76.41 & 5.86G \\ 
\hline 
\end{tabular}
}
\label{shallow-attention-comparison}
\end{table}
\paragraph{Channel Mix in GLSP Blocks}

In VRWKV~\cite{duan2024vision}, a Channel Mix module is used after the Spatial Mix. In contrast, our RWKV-UNet's encoder designs GLSP blocks without Channel Mix, as the pointwise convolution in the output layer performs both channel mixing and feed-forward operations. We compare results on the Synapse dataset (without pre-training, 100 epochs) . Table \ref{channelmix-dw-comparison} demonstrates that Channel Mix is not essential for maintaining performance.

\begin{table}[htp]
\caption{Comparison of results on the Synapse dataset: effects of Channel Mix after Spatial Mix in GLSP blocks}
\centering
\resizebox{\linewidth}{!}{
\fontsize{5pt}{6pt}\selectfont 
\begin{tabular}{cccc}
\hline 
\multirow{2}{*}{Channel Mix}  & \multicolumn{2}{c}{Average}  & \multirow{2}{*}{FLOPs} \\
\cline{2-3} 
& HD95 $\downarrow$ & DSC $\uparrow$ & \\ 
\hline 
w/& 35.19 & 74.41& 19.43G \\ 
w/o (ours) &\textbf{25.01}&  \textbf{78.14}& 11.11G \\ 
\hline 
\end{tabular}
}
\label{channelmix-dw-comparison}
\end{table}

\label{sec:channelmix-dw}

\paragraph{Skip Connections}
Comparison results of different skip connection designs for RWKV-UNet on the Synapse dataset are shown in Table \ref{skip}. Experiments analyze variations in the number of skip connections and the inclusion of the CCM block. We prioritize retaining shallow skip connections because shallow layers tend to experience greater information loss. The results show that, like other UNet's variants, adequate skip connections are essential for RWKV-UNet's performance. 
Furthermore, the results show that the CCM module can significantly improve the model segmentation results. With only two skips, the model can still achieve 84.28 DSC with CCM. Admittedly, the CCM module increases the computational load to some extent due to the aggregation of global information.

\begin{table}[htp]
\caption{Comparison results on different skip connections designs of RWKV-UNet on the Synapse dataset. Experiments are conducted on the number of skips and whether or not to use the CCM block. }
\centering
\resizebox{0.92\linewidth}{!}{
\begin{tabular}{cccccc}
\hline 
\multirow{2}{*}{Skips}&\multirow{2}{*}{CCM Block} &\multicolumn{2}{c}{Average} &\multirow{2}{*}{FLOPs}\\ 
\cline{3-4} 
&& HD95 $\downarrow$ & DSC $\uparrow$ & \\ 
\hline 
0 & w  &12.72 & 69.83 & \pzo9.18 G\\ 
1 & w/o  & 10.83 & 78.27 & \pzo9.33 G\\ 
1 & w/  & \underline{10.37} & 83.19 & 10.96 G\\ 
2&w/o  & 11.32 & 80.11 &\pzo9.41 G \\ 
2&w/  & 11.62 & \underline{84.28} &11.04 G \\ 
3&w/o  & 12.56 & 82.40& \pzo9.48 G\\ 
3 (ours)&w/ (ours) & \pzo\textbf{8.85} &\textbf{84.29}& 11.11 G\\ 
\hline 
\end{tabular}
}
\label{skip}
\end{table}
\begin{table}[htp]
\caption{Comparison of different hidden rates in the ChannelMix of CCM on the Synapse dataset.}
\centering
\resizebox{0.92\linewidth}{!}{
\fontsize{5pt}{6pt}\selectfont 
\begin{tabular}{ccccc}
\hline
\multirow{2}{*}{Hidden Rate} & \multicolumn{2}{c}{Average} & \multirow{2}{*}{FLOPs} \\
\cline{2-3}
& HD95 $\downarrow$ & DSC $\uparrow$ & \\
\hline
1 & 10.47 & 82.94 & 10.57 G \\
2 (ours) & \textbf{8.85} & \textbf{84.29} & 11.11 G \\
3 & 9.89 & 84.18 & 11.62 G \\
4 & 9.24 & 84.06 & 12.14 G \\
\hline
\end{tabular}
}
\label{hiddenrate}
\end{table}
Based on the results in Table \ref{hiddenrate}, the hidden rate of the Channel Mix layer in the CCM module materially affects performance: increasing the rate from 1 to 2 yields a notable improvement in performance, while further increases bring no gains at the cost of higher FLOPs. 
\paragraph{Larger Resolution Inputs} 
The comparison results of different methods with 512$\times$ 512 input in the Synapse dataset are shown in Table \ref{512}. The average Dice scores of other methods are from~\cite{chen20233d}. Our results indicate that increasing the input resolution improves the performance of our model. Furthermore, compared to TransUNet, the computational overhead associated with the resolution increase is less severe, showcasing the efficiency of at higher resolutions.
\begin{table}[htp]
\caption{Comparison of different methods with 512$\times$512 resolution input on the Synapse dataset. The evaluation
metric is DSC in (\%). The average DSCs of CNN-based methods are from~\cite{chen20233d}.}
\centering
\resizebox{0.92\linewidth}{!}{
\fontsize{5pt}{6pt}\selectfont 
\begin{tabular}{lcc} 
\hline
Method & Average &FLOPs \\
\hline
U-Net~\cite{ronneberger2015u} & 81.34 &- \\
Pyramid Attention~\cite{li2018pyramid} & 80.08 &- \\
DeepLabv3+~\cite{chen2018encoder} & 82.50&- \\
UNet++~\cite{zhou2018unet++} & 81.60&- \\
Attention U-Net~\cite{oktay2018attention} & 80.88&- \\
nnU-Net~\cite{isensee2021nnu}  & 82.92 &-\\
TransUNet~\cite{chen2021transunet} & \underline{84.36} &148.29G\\
SAMed\_h~\cite{samed} & 84.30&783.98G\\
RWKV-UNet (ours) & \textbf{86.73} & \pzo58.05G\\
\hline
\end{tabular}
}
\label{512}
\end{table}

%% file: secs/5_conclusion.tex
\section{Conclusion} \label{sec:conclusion}
In this study, we introduce RWKV-UNet, a novel architecture that integrates the RWKV architecture with U-Net. By combining the strengths of convolutional networks for local feature extraction and RWKV’s ability to model global context, our model significantly improves medical image segmentation accuracy. The proposed enhancements, including the GLSP module and the CCM module, contribute to a more precise representation of features and information fusion across different scales. Experimental results on 11 datasets demonstrate that RWKV-UNet surpasses SOTA methods. Its variants (RWKV-UNet-S and RWKV-UNet-T) offer a practical balance between performance and computational efficiency. Our approach has a strong potential to advance medical image analysis, particularly in clinical settings where both accuracy and efficiency are paramount.

\noindent\textbf{Limitations and Future Work.} 
RWKV-UNet is a powerful 2D medical image segmentation model that effectively combines RWKV with convolutional operations; however, it is currently not applicable to 3D imaging. In the future, we plan to extend the model to 3D for handling volume data and to explore the potential of RWKV-based foundational models for medical imaging. We also aim to develop ultra-lightweight RWKV-based models tailored for point-of-care applications, preserving segmentation accuracy while enhancing adaptability and speed further.